\documentstyle[12pt,epsfig]{article}
\title{
The  $pd\to ^3H_\Lambda K^+$ reaction cross section}
\author{
V.I. Komarov,
A.V. Lado,
\thanks{Permanent address:
 Kazakh State University,
Department of Physics, Timiryasev str. 47, 480121 Alma-Ata, Kazakh
Republic} \\  Yu.N. Uzikov
\\
\footnotesize \em Laboratory of Nuclear Problems,\\
\footnotesize \em Joint Institute for Nuclear Research, Dubna,
Moscow reg. 141980, Russia}
\begin{document}
\baselineskip=2pc

\maketitle


e-mail uzikov@nusun.jinr.dubna.su
\newpage
 \section*{Abstract}
The one- and two-step mechanisms of the $pd\to ^3H_\Lambda K^+$ reaction
in the range of incident proton kinetic energy $T_p=1.13-3.0\  GeV$ have
been investigated for the first time. A remarkable peculiarity of
the two-step mechanism  which  incorporates
 subprocesses $pp\to d\pi ^+$ and $\pi^+n\to K^+\Lambda $
 is the so called velocity matching providing the presence of all intermediate
 particles nearly to the on-mass-shell. The differential cross section
 has been
 calculated using a realistic model for the hypertritium $^3H_\Lambda $ wave
 function. The maximum value of the cross section is estimated as
$\sim $1nb/sr. The contribution of the one-step mechanism with the elementary
process $pN\to NK\Lambda $ into the cross section has been found to be
two - three
orders of magnitude smaller in comparison with the two-step mechanism.

\bibliographystyle{unsrt}
\newpage

\begin{figure}[h]
\label {fig1}
\centering
\mbox{\epsfig{figure=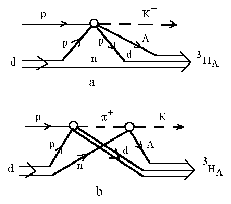,height=0.6\textheight, clip=}}
\caption{
}

\end{figure}

 The $K^+$ meson production in proton-nucleus collisions is of great interest
as these reactions allow one to investigate the nuclear structure at
short distances between nucleons \cite {cassing}.
While the experimental research programs \cite{cosy92} are supposed for the
 target nuclei with $A\geq 12$ the clearest theoretical analysis can be
 done for the
 lightest nuclei. The $pd\to ^3H_\Lambda K^+$ reaction investigated here is
 a process with high momentum transfer. So, at the threshold
of this reaction ($T_p=1132 MeV$) initial proton and deuteron have momenta
 $\sim $ 1 GeV/c in the c.m.s. but in the final state all nucleons are at
 rest. At the
proton kinetic energy in the laboratory system $T_p$ below 1580 MeV the $p+N
\to N+\Lambda +K$ process on a free nucleon N at rest is forbidden by
the energy-momentum conservation. Therefore the $pd\to ^3H_\Lambda K^+$
 reaction in this region occurs either through involving high momentum components
of the deuteron wave function
when incident proton collides  with one of its nucleons (one-step mechanism,
Fig. 1 \ref {fig1},{\it a}) or by means of active interaction with two nucleons of the
deuteron (two-step mechanism, Fig.1,\ref {fig1}{\it b}). It seems less obvious that in the last case
  the high momentum
components of the wave function will be required. In this respect the
$pd\to ^3H_\Lambda K^+$ reaction is similar to $pd\to ^3He\pi^0$ \cite
{laget1} and $pd\to ^3He \eta$ \cite {laget2} reactions for which the
two-step mechanism (called  a three -body one
in literature) was found
to dominate \cite{laget2,klu}. Indeed, the $pd\to ^3H_\Lambda+K^+$ and
$pd\to ^3He \eta$  reactions have deeper analogy in the framework of
the two-step mechanism with subprocesses $pp\to d\pi^+$  and $\pi^+n\to
\Lambda K^+$ or $\pi^+n\to p\eta$ respectively. The relation between
masses of initial
and final particles in these reactions is such that at the corresponding
threshold of the reaction as well as for the angles $\theta _{c.m.}\sim 90^o$
which determines the direction of the final meson momentum in respect to
the incident
beam, all intermediate particles ($\pi-$meson, deuteron, nucleon) are near to
on-mass-shell in a very wide energy range above the threshold \cite {kilian}.
For this reason the two-step mechanism corresponding to the Feynman graph in
 Fig.1\ref {fig1},{\it b} seems to be the most realistic model of this reaction.
It should be noted that for  production of $\pi-$mesons and heavier
mesons  ($\omega, \phi, \eta'$) as well as for target-nuclei with $A\geq 3$
 the above mentioned velocity matching does not take the place.

     Another interesting aspect of the $pd\to ^3H_\Lambda K^+$ reaction
is connected with formation of the hypertritium nucleus $^3H_\Lambda$ in the
final state. The $^3H_\Lambda$ nucleus is a loosely bound system with the
binding energy  $\varepsilon \sim
2.35 MeV$ which probably has a configuration of the $^3H_\Lambda \to
 d+\Lambda$ \cite {6}.
 An investigation of the $pd\to ^3H_\Lambda K^+$ reaction can
give a new independent information about the wave function of the $^3H_\Lambda$
nucleus.


 In the framework of the two-step mechanism the amplitude
$A^{twost}(pd\to ^3H_\Lambda K^+)$
of the $pd\to ^3H_\Lambda K^+$ reaction can be written in the full analogy
with the amplitude of the $pd\to ^3He \eta $ reaction \cite {klu}. As a result,
 we get
\begin{equation}
 A^{twost}(pd\to ^3H_\Lambda K^+)= C{\sqrt 3\over{ 2m}} A_1(pp\to d\pi^+)
 A_2(\pi^+n\to K^+\Lambda) {\cal F}(P_0,E_0)
\label{1}
\end{equation}
 where $A_1$ and $A_2$ are the amplitudes of the processes $pp\to d\pi^+$
and $\pi^+n\to K^+\Lambda$ respectively, $m$ is the nucleon mass,
$C=3/2$ is the isotopic spin factor allowing for the summation over isotopic
spin indices in the intermediate state; the nuclear formfactor in exp.
(\ref{1}) is defined as
\begin{equation}
\label{2}
{\cal F} (P_0,E_0)=\int {d^3q_1\over {(2\pi )^3}} {d^3q_2\over {(2\pi )^3}}
{\Psi_d({\bf q}_1) \Psi_H({\bf q}_2)\over {E_0^2-({\bf P}_0+{\bf q}_1+
{\bf q}_2)^2+i\epsilon}}.
\end{equation}

 Here $\Psi_d({\bf q}_1)$ is the wave function of the deuteron and
$ \Psi_H({\bf q}_2)$ is the wave function of the $^3H_\Lambda$
nucleus in the $^3H_\Lambda \to d+\Lambda$-channel in momentum space;
$E_0$ and ${\bf P}_0$ are the energy and momentum of the intermediate $\pi -$
meson at zero momenta of nucleons in the nuclear vertices ${\bf q}_1=
{\bf q}_2=0$:
\begin{equation}
 E_0=E_K+{1\over 3}E_H-{1\over 2}E_d, \ \
{\bf P}_0={2\over 3}{\bf P}_H + {1\over 2}{\bf P}_d,
\label {3}
\end{equation}
where $E_j$ is the energy of the jth particle in the c.m.s., ${\bf P}_d$
and ${\bf P}_H$ are the  momenta in the initial deuteron  and the $^3H_\Lambda$
nucleus in the c.m.s. respectively.

 According to the paper \cite {klu}, when deriving exp. (\ref {2}) we neglect
zero components $q_{10}$ and $q_{20}$  of the 4-momenta $q_{1}$ and $q_{2}$
 in the 4-dimensional propagator of $\pi-$meson
$$(p_\pi ^2 -m_\pi ^2 +
i\varepsilon)^{-1}= \{(p_K +{1\over 3}P_H -{1\over 2}P_d +q_1 -q_2)^2
-m_\pi^2 +i\varepsilon\}^{-1}$$
in comparison  with the energies $E_k,\ E_H,\ E_d$. The 3-momenta
${\bf q}_1$ and ${\bf q}_2$ are taken exactly. Recently there has appeared a
calculation \cite {faldt} for the $pd \to ^3He\eta $ reaction near the
threshold in the two-step
 model which is very similar to that developed in paper \cite {klu} and used
here. The authors of paper \cite {faldt} apply the 3-dimensional diagram
 technique and instead of the  4-dimensional $\pi -$meson
propagator $(p_\pi ^2 -m_\pi ^2 +i\varepsilon)^{-1}$ they deal with the
 energy denominator
 $(\sqrt{s_{pd}}- E_\pi-E_n -E_d+i\varepsilon )^{-1}$. The linearization procedure
over Fermi momenta ${\bf q}_1$ and ${\bf q}_2$ is used  in order to
perform
integration over $d{\bf q}_1$ and $d{\bf q}_2$. Exp. (\ref{2}) for the nuclear
formfactor differs from that in paper \cite{faldt} while in the both cases
it is a rather smooth function of kinematic variables. It is obvious that the
 reasons for this difference are  different means for consideration of
relativistic effects in the two-step models \cite{klu} and \cite {faldt}.

 The amplitude (\ref {1}) is connected to the differential cross section
of the $pd\to ^3H_\Lambda K^+$ reaction by the following expression
\begin{equation}
\label {4}
{d\sigma\over {d\Omega }}={1\over{64\pi ^2}}{1\over s_{pd}}
{|{\bf P}_H|\over {|{\bf P}_d|}} {\overline {|A(pd\to ^3H_\Lambda K^+)|^2}},
\end{equation}
where $s_{pd}$ is the invariant mass of the initial p+d state. The amplitudes
$A_1(pp\to d\pi^+)$ and $ A_2(\pi^+n\to \Lambda K^+) $ are related to
the corresponding differential cross sections by analogous relations. One
should note that the amplitudes $A_1$ and $A_2$ are factored outside the
integral sign at the point ${\bf q}_1={\bf q}_2=0$. As  mentioned in
paper \cite{faldt}, factorisation of the
 $pd\to ^3He X$ cross section in the product of $pp\to d\pi^+$
and $\pi ^+n\to\eta p$ cross sections takes  place if only one of two
 invariant forward $pp\to d\pi^+$ amplitudes dominates. For simplicity  we
assume here that this condition is fulfilled.

   The amplitude of the one-step mechanism corresponding to the Feynman graph
in Fig.1,\ref {fig1} {\it a} can be written as
\begin{equation}
\label {5}
A^{onest}(pd\to ^3H_\Lambda K^+)=\sqrt{3\over m}A_3(pN\to N\Lambda K^+)\Phi
 ({\bf} Q),
\end{equation}
where $A_3$ is the $pN\to N\Lambda K^+$ process amplitude which is factored
outside
the two-loop integration sign. The nuclear formfactor $\Phi ({\bf Q})$
is defined by
\begin{equation}
\label{6}
\Phi ({\bf }Q)= \int d^3r\varphi_d({\bf }r) \varphi_d^+({\bf } r)
\Psi_H^+({1\over 2}\bf r)\exp{(i{\bf Q} {\bf }r)},
\end{equation}
where
\begin{equation}
{\bf Q}= {1\over 3}{\bf P}_H-{1\over 2}{\bf P}_d.
\end{equation}
One should note that integral (\ref {6}) has a meaning of the deuteron
elastic formfactor $F_d(2{\bf} Q)$ at the transferred momentum $\Delta =2Q$
modified by the presence of the hypertritium
 wave function $\Psi_H({1\over 2}{\bf }r)$ in the integrand. It is obvious
that the formfactor $\Phi ({\bf} Q)$  decreases fast with  growing $Q$.


   The one-step amplitude has been numericaly calculated here
using both $S-$ and $D-$components of the deuteron wave function for the
 RSC potential in
parametrisation \cite {alberi}.
Using the experimental data on the total cross section $\sigma _{NN\to K^+
\Lambda N}$
\cite {zwerman} we estimated here the squared amplitude
$|A_3(pN\to N\Lambda K^+)|^2$ as $\sim 250 \div 450
GeV^{-2}$ in the initial proton energy range $1.6-3.0\ GeV$. The numerical
 calculations for the two-step mechanism are performed in the s-wave approximation
 for the deuteron wave function \cite{alberi}. ( As was shown by our
 calculations, the contribution  of the deuteron D-component  to the cross
section is about 10 \%). For the wave function of the $^3H_\Lambda$ nucleus
the $d+p$-model developed in Ref. \cite {6} on the basis of separable
$\Lambda N-$interaction is used. In this model the $^3H_\Lambda$ wave
function  only contains the S-component . In the S-wave approximation the
factor (\ref {2}) takes the form
\begin{equation}
{\cal F}_{000}(P_0,E_0)={1\over 4\pi}\int _0^\infty j_0(P_0r)\exp{(iE_0r)}
\varphi _d(r)\varphi_H(r)r\ dr.
\label{8}
\end{equation}
For the differential cross section of the reaction $pp\to d\pi^+$ the
parametrisation of Ref. \cite{ritchie} is used here. For the $\pi^+n\to
\Lambda K^+$ differential cross section the parametrisation of the total
cross section from Ref. \cite {9} is used and isotropic behaviour of the
cross section is assumed.

   We have investigated here numericaly the behaviour of the formfactor
${\cal F}_{000}(P_0,E_0)$ as a function of incident proton kinetic energy
$T_p$ at different $K^+$-meson scattering angles $\theta _{c.m.}$.
 The momentum $P_0$ is a rather fast decreasing function of $T_p$
at  $\theta_{c.m.}=180^o$ ($P_0=0.5-0.1 GeV/c$ in the range $T_p=
1.1-3.0 GeV $). On the contrary, at the scattering angles
$\theta_{c.m.}=0^o$ and $90^o$ both the energy $E_0$ and momentum $P_0$ are
increasing functions of $T_p$ ($E_0,P_0\sim 0.5-1.2 GeV$).
This  behaviour of $P_0$ results in a
large value of the formfactor $|{\cal F}_{000}(P_0,E_0)|^2 $ at
$\theta_{c.m.}=180^o$ in comparison to the ones at $\theta_{c.m.}=0^o$
and $90^0$. If one substitutes the wave function of the $^3He$
 nucleus  in the $d+p-$ channel \cite {zuy} instead of the $^3H_\Lambda$
hypernucleus in exp.
(\ref{8}) then the squared formfactor $|{\cal F}_{000}(P_0,E_0)|^2 $
corresponds to the one for the $pd\to ^3He\eta$ reaction and it turns out to
decrease faster  with growing incident energy $T_p$ and its value at the
threshold increases by a factor of 3 - 5.

\begin{figure}[h]
\label {fig3}
\centering
\mbox{\epsfig{figure=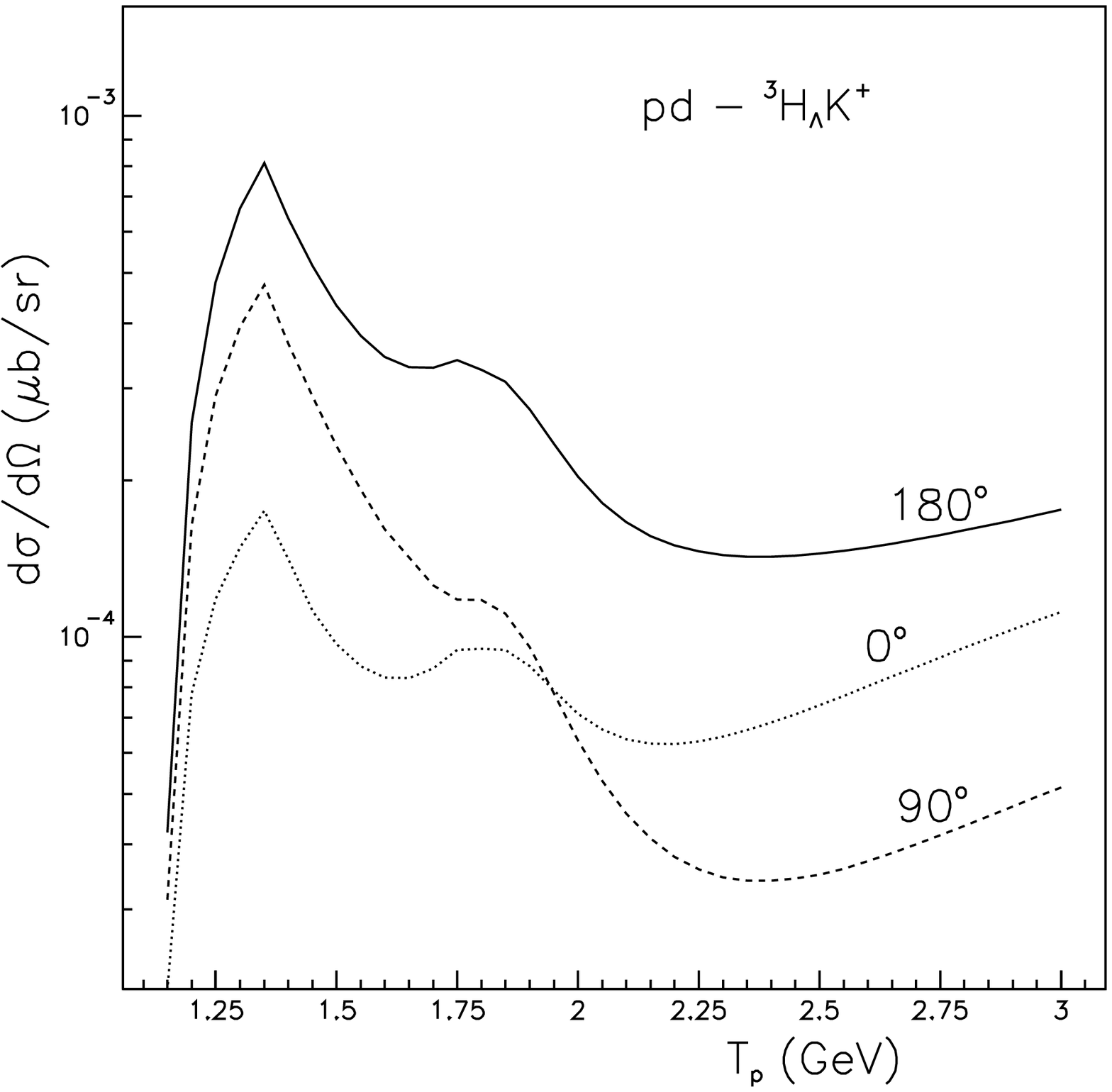,height=0.8\textheight, clip=}}
\caption{
}

\end{figure}

The calculated differential cross sections of the
$pd\to ^3H_\Lambda K^+$ reaction are presented in Fig.2 \ref {fig3}. One can see
from this picture that for any scattering angle the differential cross
section has
a sharp maximum at the proton energy $T_p\sim 1.2GeV$, which displays
the corresponding sharp peak observed in the total cross section of the
$\pi^+N\to \Lambda +K^+$ reaction (see Ref. \cite {9} and references therein).
On the whole, the relations between differential cross sections at the angles
$\theta_{c.m.}=0^o,90^o$ and $180^o$ follow from corresponding relations
between formfactors $|{\cal F}_{000}(P_0,E_0)|^2 $.

 The differential cross section of the $pd\to ^3H_\Lambda K^+$ reaction
predicted by the two-step model differs from that for the
$pd\to ^3He\eta$ reaction in two respects \cite {klu}. First, the maximum value
of the $K^+$-meson production cross section $\sim 1nb/sr$ is about 50 times
smaller than that for the $\eta-$meson production. Secondly, the
$pd\to ^3H_\Lambda K^+$ reaction cross section is a smoother decreasing
function of incident proton energy in comparison with the cross section
of the $pd\to ^3He\eta$ reaction. As follows from the behaviour of the
formfactor $|{\cal F}_{000}(P_0,E_0)|^2 $   both these
peculiarities are in part connected to the form of the wave function
of the $^3H_\Lambda$ nucleus.

\begin{figure}[h]
\label {fig5}
\centering
\mbox{\epsfig{figure=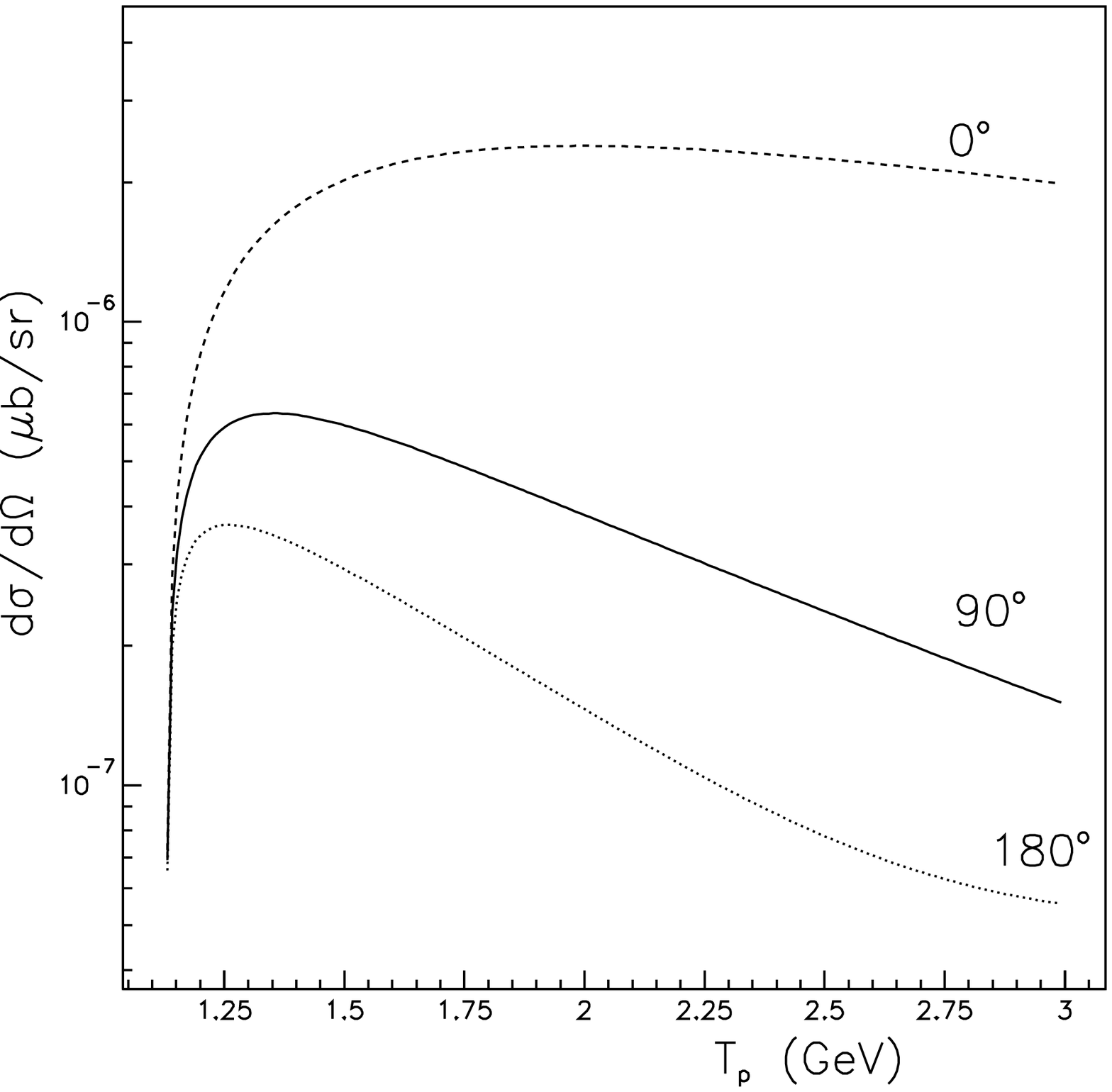,height=0.8\textheight, clip=}}
\caption{
}
\end{figure}

    The results of calculation in the framework of the one-step mechanism
are presented in Fig.3 \ref {fig5}. One can see that the contribution of this mechanism
is  two - three orders of magnitude smaller than that following from the
two-step model.

     In conclusion, we note that the two-step mechanism of
the $pd\to \ ^3H_\Lambda K^+$ reaction is used  owing to the velocity matching.
 In the case of $\eta-$meson production this mechanism
explains qualitatively the energy dependence of the cross section above the
threshold \cite {klu}. However, just at the threshold this model is in strong
contradiction with the experimental data on the $pd\to ^3He\eta$ reaction
\cite {klu}.
One of a reason for it is probably a strong attractive interaction in the
final $\eta-^3He$ state caused by  an excitation of the nucleon $N^*(1535)$
resonance \cite{klu, wilk}. At present there are no experimental data
pointing to the presence of strong coupling of the $K^+-$meson to any nucleon
resonance in the resonance mass region of $1.2- 2.0\ GeV$. Therefore one can
 suppose that final state interaction in the $pd\to ^3H_\Lambda K^+$ reaction
will  not be of great importance in contrast to the $\eta-$production.
\newpage

Authors are sincerely grateful to A.V. Kondratyuk  and L. Mailing for
useful discussion. This work was supported in part by grant $N^o$
93-02-3745  of the Russian Foundation for Fundamental Researches.

\eject

\newpage
\section*{Figure captions}
 Fig.1 The one- step ({\it a}) and two-step ({\it b}) mechanisms
of the $pd\to ^3H_\Lambda K^+$ reaction.
\newline
Fig.2.
The differential cross section of the $pd\to ^3H_\Lambda K^+$ reaction
 calculated for the two-step mechanism as a function of incident proton kinetic
energy at different angles of $K^+$-meson
 $\theta _{c.m.}=0^o,\ 90^o,\ 180^o$
\newline
Fig.3.
The same as in Fig.3 but for the one-step mechanism

\newpage \begin{thebibliography}{99}
\bibitem{cassing}  Cassing W, Batko G, Mosel U, Niita K, Shult O,
Wolf Gy. 1990 {\it  Phys. Lett.} {\bf B 238}
 (1990) 25; Sibirtsev A, B\"usher M. 1994 {\it Z. Phys.} {\bf A347}  191;
 Kacharava A, Macharashvili G,  Mamulashvili A,
Menteshashvili Z, Nioradze M, Komarov V I. 1994 {\it HEPI TSU} 12-14.
\bibitem{cosy92} Sistemich K. et al. 1992 {\it COSY Proposal} $N^o$ 18.
\bibitem{laget1} Laget J M, Lecolley J F. 1987 {\it Phys. Lett.} {\bf B 194} 177.
\bibitem{laget2} Laget J M, Lecolley J F. 1988 {\it Phys. Rev. Lett.}
 {\bf 61} 2069.
\bibitem{klu} Kondratyuk L A, Lado A V,  Uzikov Yu N. 1995 {\it Yad.Fiz.}
{\bf 57} 524.
\bibitem{faldt} F\"aldt G , Wilkin C. 1995 {\it Nucl. Phys.} {\bf A587} 769.
\bibitem{kilian} Kilian K, Nann H. Preprint KFA, Juelich (1989).
\bibitem{6} Congleton J G. 1992 {\it J. Phys.G: Nucl. Part.} {\bf 18} 339.
\bibitem{alberi} Alberi G,  Rosa L P, Thome Z D. 1975 {\it Phys. Rev. Lett. }
{\bf 34} 503.
\bibitem{zwerman} Zwerman W. 1988 {\it Mod. Phys. Let.} {\bf A 3} 251.
\bibitem{ritchie} Ritchie B G. 1991 {\it Phys.Rev.} {\bf C44} 533.
\bibitem{9} Cugnon J, Lombard R M. 1984 {\it Nucl. Phys.} {\bf A 422} 635.
\bibitem{zuy} Zhusupov M A, Uzikov Yu N, Yuldasheva G A. 1986 {\it Izv. AN
KazSSR, ser.fiz.-mat.} {\bf  6}  69.

\bibitem{wilk} Wilkin C. 1993 {\it Phys. Rev.} {\bf C47} R938.


\end{thebibliography}
\end{document}